Technical Note

# Multi-Fidelity Surrogate Based on Single Linear Regression

Yiming Zhang, Nam-Ho Kim, Chanyoung Park, Raphael T. Haftka
Department of Mechanical and Aerospace Engineering
University of Florida
Gainesville, Florida, 32611
{yimingzhang521, nkim, cy.park, haftka}@ufl.edu

**Introduction**

Various frameworks have been proposed to predict mechanical system responses by combining data from different fidelities for design optimization and uncertainty quantification as reviewed by Fernández-Godino et al.[1] and Peherstorfer et al. [2]. Among all frameworks, the Bayesian framework based on Gaussian processes [3] has the potential of highest accuracy. However, the Bayesian framework requires optimization for estimating hyper-parameters, and there is a risk of estimating inappropriate hyper-parameters as Kriging surrogate often does, especially in the presence of noisy data. We propose an easy and yet powerful framework for practical design and applications. In this technical note, we revised a heuristic framework [4] which minimizes the prediction errors at high-fidelity samples using optimization. The system behavior (high-fidelity behavior) is approximated by a linear combination of the low-fidelity predictions and a polynomial-based discrepancy function. The key idea is to consider the low-fidelity model as a basis function in the multi-fidelity model with the scale factor as a regression coefficient. The design matrix for least-square estimation consists of both the low-fidelity model and discrepancy function. Then the scale factor and coefficients of the basis functions are obtained simultaneously using linear regression, which guarantees the uniqueness of fitting process. Besides enabling efficient estimation of the parameters, the proposed least-squares multi-fidelity surrogate (LS-MFS) can be applicable to other regression models by simply replacing the design matrix. Therefore, the LS-MFS is expected to be easily applied to various applications such as prediction variance, D-optimal designs, uncertainty propagation [6, 7] and design optimization.

**Least-squares Multi-fidelity Surrogate**

A heuristic multi-fidelity surrogate (MFS) using polynomial response surface (PRS) has been proposed[4] and demonstrated reasonable effectiveness and robustness. The MFS provides a fit combining a small number of expensive high-fidelity data and less expensive low-fidelity data as a function of design parameters. The MFS is built with two surrogates, $\hat{f}_L(\mathbf{x})$ and $\hat{\delta}(\mathbf{x})$, which are the PRS fitted to the low fidelity data and the discrepancy data as

$$\hat{f}_H(\mathbf{x}) = \rho \hat{f}_L(\mathbf{x}) + \hat{\delta}(\mathbf{x}) \tag{1}$$

where the scale factor $\rho$ and discrepancy function $\hat{\delta}(\mathbf{x})$ are obtained through optimization as

$$\min_{\rho, \hat{\delta}(\mathbf{x})} : \left(\hat{\delta}(\mathbf{x}_H) - \mathbf{d}_H\right)^T \left(\hat{\delta}(\mathbf{x}_H) - \mathbf{d}_H\right) \tag{2}$$

$$\mathbf{d}_H = \rho \hat{f}_L(\mathbf{x}_H) - \mathbf{y}_H \tag{3}$$

where the high-fidelity data set is denoted as ($\mathbf{x}_H$, $\mathbf{y}_H$) including $n$ samples.

In this technical note, we propose the least-squares estimation of the scale factor $\rho$ and the coefficients of the discrepancy function. The MFS in Eq. (1) can be expanded with the discrepancy function being represented by $p$ monomial bases, as

$$\hat{f}_H(\mathbf{x}) = \rho \hat{f}_L(\mathbf{x}) + \hat{\delta}(\mathbf{x})$$
$$= \rho \hat{f}_L(\mathbf{x}) + \sum_{i=1}^{p} X_i(\mathbf{x}) b_i \qquad (4)$$

where $X_i(\mathbf{x})$ denotes the $i^{th}$ monomial/basis, and $b_i$ is the coefficient of $X_i(\mathbf{x})$. In the proposed LS-MFS, the traditional design matrix is augmented by including the low-fidelity model with unknown scale parameter $\rho$. In least-squares estimation, the relationship between high-fidelity samples and model predictions can be given as

$$\mathbf{Y} = \mathbf{XB} + \mathbf{e} \qquad (5)$$

where

$$\mathbf{Y}_{n\times 1} = \begin{bmatrix} \mathbf{y}_H^{(1)} \\ \vdots \\ \mathbf{y}_H^{(n)} \end{bmatrix}, \mathbf{X}_{n\times(p+1)} = \begin{bmatrix} \hat{f}_L(\mathbf{x}_H^{(1)}) & X_1(\mathbf{x}_H^{(1)}) & \cdots & X_p(\mathbf{x}_H^{(1)}) \\ \vdots & \vdots & \ddots & \vdots \\ \hat{f}_L(\mathbf{x}_H^{(n)}) & X_1(\mathbf{x}_H^{(n)}) & \cdots & X_p(\mathbf{x}_H^{(n)}) \end{bmatrix}, \mathbf{B}_{p+1} = \begin{bmatrix} \rho \\ b_1 \\ \vdots \\ b_p \end{bmatrix}, \mathbf{e}_{n\times 1} = \begin{bmatrix} \mathbf{e}_H^{(1)} \\ \vdots \\ \mathbf{e}_H^{(n)} \end{bmatrix} \qquad (6)$$

In the above equation, $\mathbf{Y}$ is the vector of high-fidelity samples, $\mathbf{X}$ is the augmented design matrix, $\mathbf{B}$ is the parameter vector and $\mathbf{e}$ is the vector for residual errors. By augmenting the design matrix, the scale parameter and unknown coefficients of discrepancy functions are estimated simultaneously. In addition, the low-fidelity model is considered as additional basis function. An additional advantage of the proposed LS-MFS is that it is unnecessary to build the surrogate model of the low-fidelity. It only requires to evaluate the low-fidelity model at the same locations with the high-fidelity samples. The unknown parameters in LS-MFS are obtained by using the standard regression technique as

$$\mathbf{B} = (\mathbf{X}^T \mathbf{X})^{-1} \mathbf{X}^T \mathbf{Y} \qquad (7)$$

The scale factor $\rho$ implies the level of trend similarity between $\hat{f}_H(\mathbf{x})$ and $\hat{f}_L(\mathbf{x})$, and plays a critical role to approximate multi-fidelity data. Negative values or extremely large values of $\rho$ indicates a risky prediction, which are likely to be associated with undesirable low-fidelity models, inappropriate surrogate forms, or inadequate samples. The discrepancy function $\hat{\delta}(\mathbf{x})$ is likely to be a low-order PRS when $f_L(\mathbf{x})$ had a similar trend as $f_H(\mathbf{x})$.

There are nice properties of LS-MFS. Firstly, the LS-MFS could be easily applied to more than two-fidelity models by augmenting the design matrix with multiple low-fidelity models with multiple scale factors, although we showed the LS-MFS for two-fidelity models in this technical note. Secondly, comparing with the heuristic approach, the LS-MFS is more convenient for analytical study of various applications such as the prediction variance, D-optimal design of experiments, uncertainty propagation, and design optimization, while incorporating the effect of the multi-fidelity models.

**Numerical performance of the least-squares multi-fidelity surrogate**
We selected an exponential function [8, 9] with two variables to test the LS-MFS. The high-fidelity model $f_H(\mathbf{x})$ is given in Eq. (8). A synthetic noise following a normal distribution $N(0, 0.2^2)$ has been added to the high-fidelity samples. We also modified the original low-fidelity model [10] with a larger scale factor and added a quadratic function as in Eq. (9). These variations make the multi-fidelity modeling more challenging. Major settings of the test function are summarized in Table 1. The responses of $f_H(\mathbf{x})$ (no noise) and $f_L(\mathbf{x})$ have reasonable spatial complexity as shown in Fig. 1.

$$f_H(\mathbf{x}) = \left[1 - \exp\left(-\frac{1}{2x_2}\right)\right] \frac{2300x_1^3 + 1900x_1^2 + 2092x_1 + 60}{100x_1^3 + 500x_1^2 + 4x_1 + 20} \quad (8)$$

$$f_L(\mathbf{x}) = \frac{1}{8}[f(x_1 + 0.05, x_2 + 0.05) + f(x_1 + 0.05, \max(0, x_2 - 0.05))] +$$
$$\frac{1}{8}[f(x_1 - 0.05, x_2 + 0.05) + f(x_1 - 0.05, \max(0, x_2 - 0.05))] + \quad (9)$$
$$\frac{1}{8}(-5x_1 - 7x_2^2)$$

Table 1. Major settings for the exponential test function

| Input variables | Range of $f_H(\mathbf{x})$ | Range of $f_L(\mathbf{x})$ | Noise for $f_H(\mathbf{x})$ |
|---|---|---|---|
| $x_1, x_2 \in [0,1]$ | [1.1804, 13.7692] | [-0.1867, 4.9498] | $N(0, 0.2^2)$ |

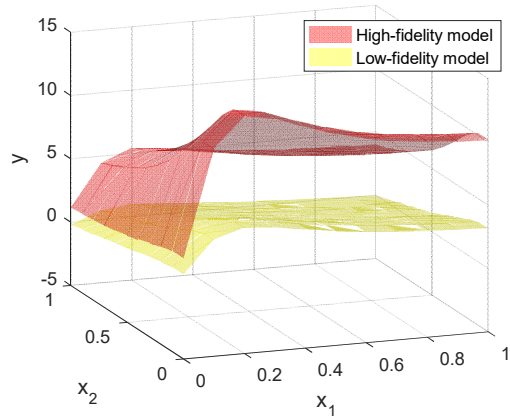

Figure 1 . Responses of the test function in 2D space

We generated two sets of samples according to $2 \times 2$ and $3 \times 3$ full factorial designs as shown in Fig. 2(a) and Fig. 2(b). $\mathbf{y}_H$ and $\mathbf{y}_L$ were computed from $f_H(\mathbf{x})$ and $f_L(\mathbf{x})$, respectively, at the full factorial designs. Instead of the low-fidelity surrogate $\hat{f}_L(\mathbf{x})$, the exact low-fidelity model $f_L(\mathbf{x})$ is used to avoid approximation error in building the low-fidelity surrogate. It is also possible to build $\hat{f}_L(\mathbf{x})$ based on only low-fidelity data for repeated calls of LS-MFS in practical applications. The LS-MFS was evaluated using relative root-mean-square error (RMSE) for the overall accuracy and maximum error for the worst individual prediction. The relative RMSE and maximum error were obtained based on $11 \times 11$ test grids as shown in Fig. 2(c).

The key parameters and performance of LS-MFS were summarized in Table 2. For the $2 \times 2$ full factorial design, LS-MFS had the relatively small RMSE of 9.00% when approximating the complex test function with noise using a linear discrepancy function. The large maximum relative error, 31.37%, was mainly due to a small function value. While introducing more samples using the $3 \times 3$ design, all the evaluation metrics (e.g. the relative RMSE, the maximum relative/absolute errors) decreased noticeably as in Table 2. PRS fitting for only high-fidelity samples were given in Table 3 as a comparison. For the $2 \times 2$ design, relative RMSE of PRS fitting was 5 times larger than that of LS-MFS, and the absolute maximum error of PRS fitting was 10 times larger than that of LS-MFS. For the $3\times 3$ design, PRS fitting improved while introducing more samples, but was still significantly inferior to the LS-MFS regarding the specified evaluation metrics.

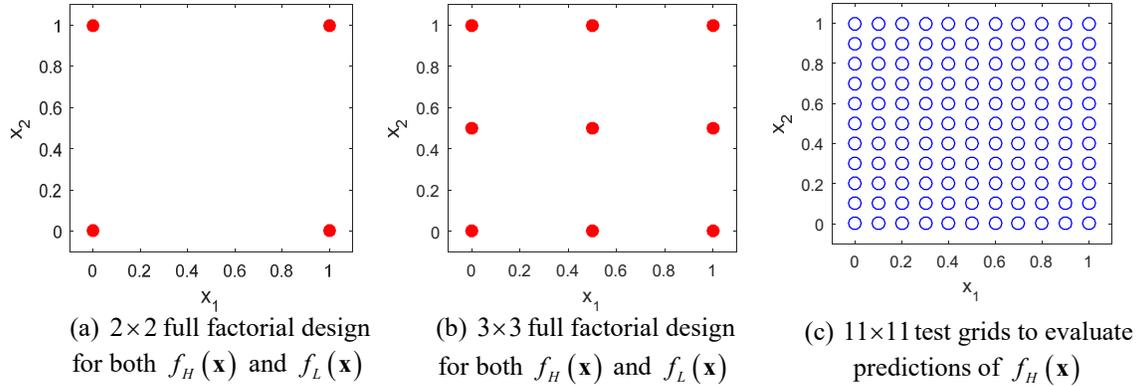

(a) $2\times 2$ full factorial design for both $f_H(\mathbf{x})$ and $f_L(\mathbf{x})$

(b) $3\times 3$ full factorial design for both $f_H(\mathbf{x})$ and $f_L(\mathbf{x})$

(c) $11\times 11$ test grids to evaluate predictions of $f_H(\mathbf{x})$

Figure 2. Design of experiments to build and test the least squares multi-fidelity surrogate (LS-MFS)

Table 2. LS-MFS based on $2\times 2$ and $3\times 3$ full factorial designs

|  | Model parameters for LS-MFS | Relative RMSE (%) | Maximum relative error (%) | Maximum absolute error |
|---|---|---|---|---|
| $2\times 2$ full factorial design | $\rho$=3.40, Linear discrepancy | 9.00% | 31.37% | 0.82 |
| $3\times 3$ full factorial design | $\rho$=3.16, Quadratic discrepancy | 1.52% | 8.03% | 0.26 |

Table 3. PRS fitting for only high-fidelity samples based on $2\times 2$ and $3\times 3$ full factorial designs

|  | Order of PRS | Relative RMSE (%) | Maximum relative error (%) | Maximum absolute error |
|---|---|---|---|---|
| $2\times 2$ full factorial design | Linear | 43.80% | 92.23% | 8.99 |
| $3\times 3$ full factorial design | Quadratic | 26.18% | 56.21% | 6.99 |

**Acknowledgments**
This work was supported by the U.S. Department of Energy, National Nuclear Security Administration, Advanced Simulation and Computing Program, as a Cooperative Agreement under the Predictive Science Academic Alliance Program, under Contract No. DE-NA0002378